\documentclass[12pt,notoc]{JHEP3}
\pdfoutput=1
\usepackage{amsmath,amssymb,euscript,array,mathrsfs}
\usepackage{epsfig}
\usepackage{graphicx}

\def\Tr{{\rm Tr}}

\def\R{\boldsymbol{R}}

\def\R{R_{\rm AdS}}

\newcommand{\Dslash}{D\mkern-11.5mu/\,} 
\newcommand{\delslash}{{\partial\mkern-9mu/}}

\def\Dbarslash{\,\,{\raise.15ex\hbox{/}\mkern-12mu {\bar D}}}
\def\Dslash{\,\,{\raise.15ex\hbox{/}\mkern-12mu D}}
\def\delslash{\,\,{\raise.15ex\hbox{/}\mkern-9mu \partial}}
\def\delbarslash{\,\,{\raise.15ex\hbox{/}\mkern-9mu {\bar\partial}}}

\def\a{\alpha}
\def\b{\beta}

\def\d{\delta}
\def\e{\epsilon}

\def\m{\mu}
\def\n{\nu}

\def\p{\pi}

\def\t{\tau}

\def\x{\chi}

\def\G{\Gamma}
\def\D{\Delta}

\newcommand{\EQ}[1]{\begin{equation} #1\end{equation}}

\newcommand{\SP}[1]{\begin{equation}\begin{split} #1\end{split}\end{equation}}

\title{Vacuum Ambiguity in de Sitter Space at Strong Coupling}
\author{Jimmy A. Hutasoit \\\\
{\it Department of Physics,\\Carnegie Mellon University,\\
Pittsburgh, PA 15213, USA.
}\\
E-mail: \email{jhutasoi@andrew.cmu.edu}
} 
\abstract{It is well known that in the weak coupling regime, quantum field theories in de Sitter space do not have a unique vacuum, but a class of vacua parametrized by a complex parameter $\a$, {\it i.e.}, the so-called $\a$-vacua. In this article, using gauge/gravity duality, we calculate the symmetric two-point function of strongly coupled ${\cal N}=4$ supersymmetric Yang-Mills theory on $dS_3$. We find that there is a class of de Sitter invariant vacua, parametrized by a set of complex parameters $\{\a_{\n}\}$. } 

\begin{document}
\section{Introduction}
Understanding de Sitter space is important because de Sitter space plays a central role in cosmology, not only at the early universe, during the inflationary epoch when the seeds of cosmic structure were generated, but also in the late universe as the cosmological constant dominates over other matter contents in the universe. The ultimate goal is to understand the full quantum gravity of de Sitter space, but even at the semi-classical level, where one considers quantum field theory living in a fixed de Sitter background, many interesting features appear. Here in this article, we will focus on the vacuum structure of a quantum field theory, namely ${\cal N}=4$ supersymmetric Yang-Mills theory, on a three-dimensional de Sitter space at the strong coupling regime.

At the level of free theory, by considering the symmetric two-point function of a free massive scalar theory, it can be shown that there is an infinite family of vacua that are invariant under the isometries of de Sitter space. This is different from Minkowski space, where the symmetries of the theory determine a unique Poincar\'{e} invariant vacuum. The existence of this vacuum ambiguity in de Sitter space was first emphasized by Mottola \cite{Mottola:1984ar} and Allen \cite{Allen:1985ux}.  This class of de Sitter invariant vacua is often parametrized by a complex parameter $\a$, and thus are usually called the $\a$-vacua.

One of the $\a$-vacua, the Bunch-Davies \cite{Bunch:1978yq} vacuum, stands prominently among others as it is the only one that behaves thermally when viewed by an Unruh detector \cite{Birrell:1982ix} and reduces to the standard Minkowski vacuum when the de Sitter radius is taken to infinity. The correlators in the Bunch-Davies vacuum can be obtained by analytical continuation from the Euclidean theory, thus it is also known as the ``Euclidean" vacuum. The difference between a correlator in an $\a$-vacuum and the one in the Bunch-Davies vacuum is often thought of as arising from an image source at the antipodal point, behind the event horizon.

The early studies of a weakly interacting scalar theory in an $\a$-vacuum found that new divergences appear which, unlike in the Bunch-Davies vacuum, cannot be renormalized \cite{Danielsson:2002mb,Collins:2003zv,Collins:2003mk}. Self-energy graphs in a theory with a cubic interaction produce pinched singularities \cite{Einhorn:2002nu} or require peculiar non-local counterterms \cite{Banks:2002nv}\footnote{Beside the Bunch-Davies vacuum, Ref. \cite{Banks:2002nv} points out another special vacuum, in which the Green's function can be interpreted as living on elliptic de Sitter space \cite{Parikh:2002py}.}. However, when one modifies the generating functional of the theory, these divergences disappear \cite{Collins:2003mj} (see also \cite{Goldstein:2003ut,Goldstein:2003qf,Einhorn:2003xb}). 

With the advent of the gauge/gravity duality \cite{Maldacena:1997re,magoo}, it is only natural to ask what happens with this vacuum ambiguity as one goes to the strong coupling\footnote{For earlier efforts in trying to connect this issue to holography, see for example \cite{Danielsson:2002qh}.}. A context in which we can ask such a question is the ${\cal N}=4$ supersymmetric Yang-Mills theory living in a three-dimensional de Sitter space (times a circle). This theory is dual to the type IIB superstring theory living in time-dependent, asymptotically locally anti de Sitter (AdS) backgrounds found in \cite{Birmingham:2002st,Balasubramanian:2002am,Cai:2002mr,Ross:2004cb,Balasubramanian:2005bg}\footnote{For other types of time-dependent AdS backgrounds, see for example \cite{Cvetic:2003zy}.}. 

The first two articles, Refs. \cite{Birmingham:2002st,Balasubramanian:2002am}, consider a double analytic continuation of AdS-Schwarzschild black holes where the time coordinate is analytically continued to Euclidean signature $t\to i\chi$, with $\x$ is periodically identified, and  a polar angle is analytically continued to Lorentzian signature $\theta \to i\t$, with $\t$ is the new time coordinate. The results are smooth, time-dependent AdS backgrounds called AdS ``bubbles of nothing." Refs. \cite{Cai:2002mr,Ross:2004cb,Balasubramanian:2005bg} then realize that there is another spacetime with the same AdS asymptotics as the bubble geometries, the so-called ``topological AdS black hole." It is a quotient of AdS space obtained by an identification of global $AdS_5$ along a boost \cite{Banados:1997df, Banados:1998dc}. The topological AdS black hole can also be obtained by a Wick rotation of thermal AdS space.

These two different types of geometries are dual to two different phases of the strongly coupled, large $N$ gauge theory formulated on the $dS_3 \times S^1$ boundary. As in the more widely known thermal ${\cal N}=4$ Super Yang-Mils case, where the field theory lives on $S^3 \times S^1$, the two phases are distinguished by the expectation value of the Wilson loop around the $S^1$. In the bubble of nothing phase, the circle shrinks to zero size in the interior of the geometry and the Wilson loop is non-zero, indicating the spontaneous breaking of the ${\mathbb Z}_N$ symmetry of the gauge theory. The topological black hole phase is ${\mathbb Z}_N$ invariant. Unlike the thermal situation however, the spontaneous breaking of ${\mathbb Z}_N$ invariance is not a deconfinement transition since the circle is a spatial direction and not the thermal circle. Indeed, studying the retarded scalar glueball propagators suggests that the topological black hole phase, which is the ${\mathbb Z}_N$ invariant phase, corresponds to the ${\cal N}=4$ theory on $dS_3\times S^1$ is in a plasma-like or deconfined state in the exponentially expanding universe, while the bubble of nothing phase corresponds to the hadronized phase \cite{Hutasoit:2009xy}.

\begin{figure}[h]
\begin{center}
\includegraphics[width=3.5in]{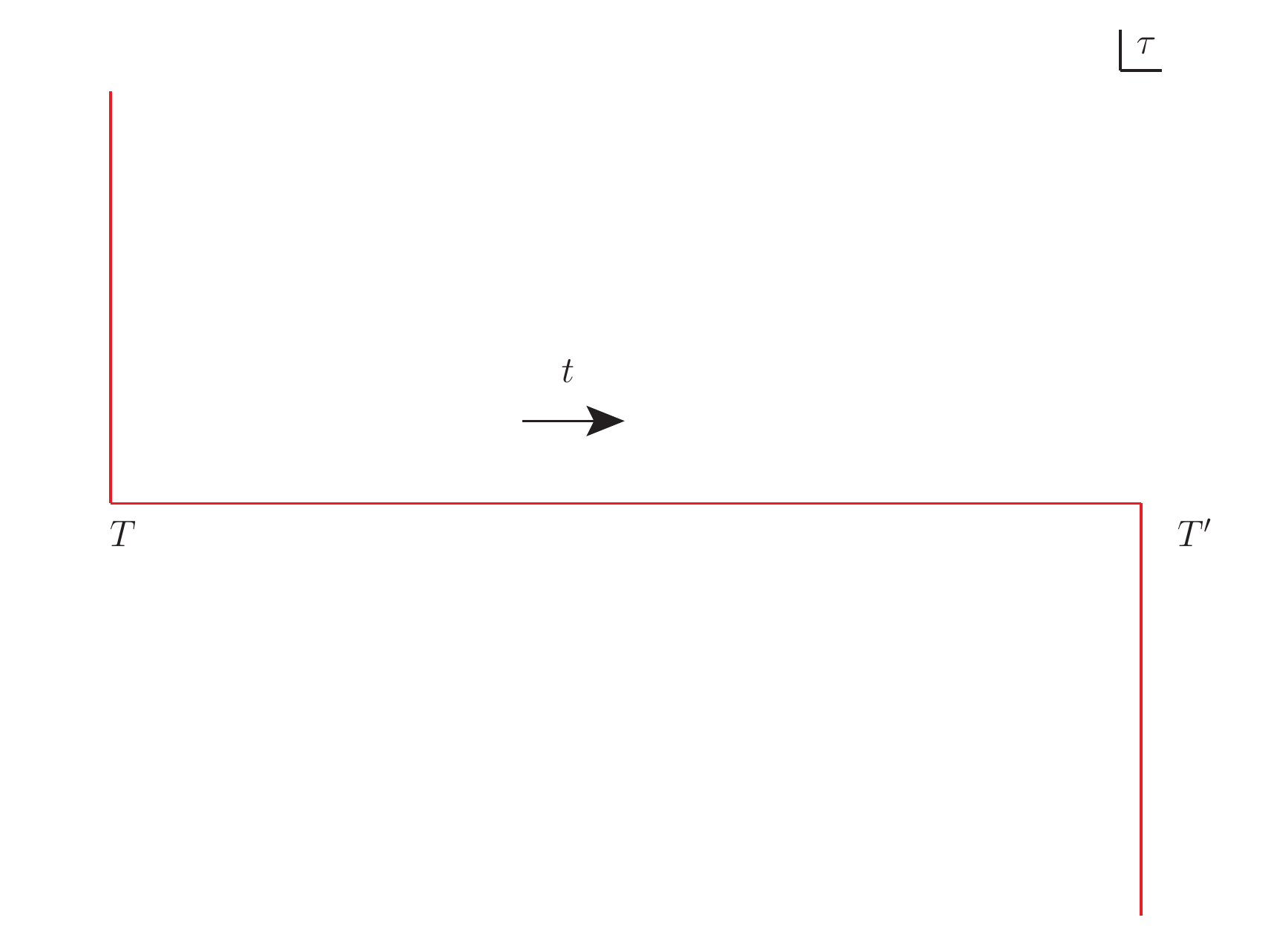} 
\end{center}
\caption{\footnotesize Son-Starinets prescription is equivalent to performing path integral on vacuum-to-vacuum contour in the complex time plane, as depicted above. The vacuum state is obtained by projection in the Euclidean path integral.}
\label{Euclid}
\end{figure}

Concerning the issue of vacuum ambiguity in de Sitter space, Ref. \cite{Hutasoit:2009xy} find that Son-Starinets prescription \cite{Son:2002sd}, which is used to calculate the retarded propagators, automatically implies that the boundary theory is in the Bunch-Davies vacuum. This is due to the fact that having the relevant boundary conditions in Son-Starinets prescription is equivalent to preparing the states in the boundary field theory by the mean of Euclidean projection \cite{vanRees:2009rw} (see Fig. \ref{Euclid}). Therefore, one can only obtain the propagators for the Euclidean vacuum using this prescription.

In this paper, using gauge/gravity duality, we calculate the symmetric two-point function of the strongly coupled ${\cal N}=4$ Super Yang-Mills as a function of the geodesic distance. Here, we will focus only on the ${\mathbb Z}_N$-invariant phase, which corresponds to the topological black hole geometry. We find that there is an ambiguity in the two-point function that arises from the vacuum ambiguity of the boundary field theory. From the bulk point of view, this ambiguity comes from the fact that there are an infinite family of bulk radial wave functions that are normalizable deep into the bulk, at the horizon of the topological AdS black hole \cite{Balasubramanian:1998de}. 

Thus, the strongly coupled ${\cal N}=4$ Super Yang-Mills on $dS_3$ has an infinite class of vacua, parametrized by a set of complex parameters $\{\a_{\n}\}$. This is different from the case of the weakly coupled theory, where the infinite family of vacua is parametrized by a single complex parameter $\a$. We do not have a satisfactory explanation for this, but one possibility is that as one increases the coupling constant of the theory, going from the free theory toward the strongly coupled one, the $\a$-vacua mix with one another. 

The article is organized as follows.  In Section \ref{alpha}, we review the issue of vacuum ambiguity in de Sitter space at the weak coupling regime. As the holographic calculation involves classical gravity calculation in the topological AdS black hole, we review some properties of the topological black hole geometry in Section \ref{toprev}. In Section \ref{calc}, we perform the holographic computation of the symmetric two-point functions as functions of the geodesic distance. We end with some discussions in Section \ref{discussions}. 

\section{$\a$-vacua in de Sitter Space \label{alpha}}
The three-dimensional de Sitter space $dS_3$ can be realized as the hypersurface described by 
\EQ
{-x_0^2+x_1^2+x_2^2+x_3^2 = R_{\rm dS}^2,
}
with $R_{\rm dS}$ being the $dS_3$ radius. The metric for the global patch is given by
\EQ
{ds^2 = -{dt^2\over R_{\rm dS}^2}+\cosh^2 \left({t\over R_{\rm dS}}\right)  \;d\Omega_2^2 \label{dsmetric}.
}
For a review on the classical geometry of de Sitter space and its other aspects see \cite{Spradlin:2001pw}. This spacetime does not contain a globally time-like Killing vector and thus does not have a unique vacuum state. To see this in a more formal manner, let us consider the symmetric two-point function in a de Sitter invariant state $|\a\rangle$. For simplicity, we consider only a free massive scalar field theory here. For the treatment of weakly interacting theories, see \cite{Collins:2003mj} and references therein.

$|\a\rangle$ being de Sitter invariant implies that the symmetric two-point function $G_{\a}(x,y)$ can only depend on the two spacetime points $x$ and $y$ via the geodesic distance $d(x,y)$. The symmetric two-point function obeys the massive scalar field equation
\EQ
{\left(-\Box_x + m^2 \right) G(x,y) = 0,
}
which can be expressed as
\EQ
{\left(Z^2-1\right)^{-1/2} \partial_Z\left[\left(Z^2-1\right)^{3/2} \, \partial_Z G\left(Z(x,y)\right) \right] + m^2\, R_{\rm dS}^2\, G\left(Z(x,y)\right)=0, \label{dsode}
}
where we have introduced the real function
\EQ
{Z(x,y) = \cos{d(x,y) \over R_{\rm dS}}\, .
}
Here, the mass term may include the contribution that comes from coupling to the curvature.

A solution is given by
\EQ
{G_E\left(Z(x,y)\right) = {\G(h_+) \, \G(h_-) \over (4 \p)^{3/2} \, \G(3/2)} \,\, _2F_1\left(h_+,h_-;{3 \over 2};{1+Z(x,y) \over 2}\right),
}
with
\EQ
{h_{\pm} = 1 \pm \sqrt{1-m^2 R_{\rm dS}^2} \,\, .
}
Here, the normalization is chosen such that the short distance singularity matches to that of Minkowski space.

We note that if $G_E(Z)$ is a solution to Eq. (\ref{dsode}), then $G_E(-Z)$ is also a solution. Since for $m^2>0$ these two solutions are linearly independent, then the general solution to Eq. (\ref{dsode}) can be expressed as
\EQ
{G_{\a}(x,y)= {1+e^{\a+\a^*} \over 1-e^{\a+\a^*}} \, G_E\left(Z(x,y)\right) + {e^{\a}+e^{\a^*} \over 1-e^{\a+\a^*}} \, G_E\left(-Z(x,y)\right). \label{alphaweak}
}
The constants in the $\a$-vacuum propagator (\ref{alphaweak}) are chosen such that the modes of the free massive scalar field $\Phi$ in the $\a$-vacuum can be related to those in the Bunch-Davies vacuum by the Bogoliubov transformation
\EQ
{\Phi_n^{\a} = {\Phi_n^E + e^{\a}\, \left(\Phi_n^E\right)^* \over \sqrt{1 - e^{\a+\a^*}}}\,.
}
The short-distance singularity of the $\a$-vacuum propagator (\ref{alphaweak})  is related to the singularity of the Bunch-Davies vacuum propagator by a factor $\left(1+e^{\a+\a^*}\right)/\left(1-e^{\a+\a^*}\right)$. 

Let us also mention a peculiar, but widely used, interpretation of the $\a$-vacuum propagator (\ref{alphaweak}). For a point $x=(t,\theta,\phi)$, we can introduce its antipodal point $\bar{x}=(-t,\theta+\p,\phi+\p)$. It is easy to show that $Z(x,\bar{y})=-Z(x,y)$. Therefore, the $\a$-vacuum propagator (\ref{alphaweak}) can be rewritten as
\SP
{G_{\a}(x,y) = &\,\,  {1 \over 1-e^{\a+\a^*}} \, G_E\left(Z(x,y)\right) + {e^{\a+\a^*} \over 1-e^{\a+\a^*}} \, G_E\left(Z(\bar{x},\bar{y})\right) \\
& + {e^{\a} \over 1-e^{\a+\a^*}} \, G_E\left(Z(\bar{x},y)\right)+ {e^{\a^*} \over 1-e^{\a+\a^*}} \, G_E\left(Z(x,\bar{y})\right). \label{antipode}
}
The interpretation is then that the extra contributions in the $\a$-vacuum propagator (\ref{alphaweak}) can be thought of as arising from an image source at the antipodal point. We would like to emphasize that even though this is an interesting interpretation, the existence of vacuum ambiguity in de Sitter space does not depend on this interpretation concerning image source at the antipodal point. 

Finally, let us remark that there is no de Sitter invariant vacuum in the case of massless minimally coupled scalar fields \cite{Allen:1985ux}. In the case of interest, since the ${\cal N}=4$ Super Yang-Mills is conformally coupled to the de Sitter background ({\it cf.} \cite{Hollowood:2006xb}), there is an infinite class of de Sitter invariant vacua at weak coupling and, as we will see in Section \ref{calc}, this vacuum ambiguity persists at strong coupling.

\section{The Topological AdS Black Hole \label{toprev}}
The so-called topological black hole in $AdS_5$ (times a five-dimensional sphere) can be obtained as a near horizon limit of $N$ D3-branes filling a boost orbifold ${\mathbb R}^{1,1}/{\mathbb Z}$ \cite{Balasubramanian:2005bg}. It is an orbifold of the $AdS_5$ space, obtained by an identification of points along the orbit of a Killing vector
\EQ
{\xi = \frac{r_\chi}{R_{\rm AdS}} \left(x_4 \partial_5 + x_5 \partial_4 \right),
}
where $r_\x$ is an arbitrary real number and the $AdS_5$ is described as the universal covering of the hypersurface
\EQ
{-x_0^2 + x_1^2 + x_2^2 + x_3^2 + x_4^2 - x_5^2 = - R^2_{\rm Ads},
}
$R_{\rm AdS}$ being the $AdS_5$ radius. In Kruskal-like coordinates which cover the whole spacetime, the metric has the form
\EQ
{
ds^2= {4R^2_{\rm AdS}\over (1-y^2)^2}\;dy^\mu dy^\nu \eta_{\mu\nu}+ 
{(1+y^2)^2\over (1-y^2)^2}\;r_\x^2d\chi^2,
\label{kruskal}
}
where $\chi$ is a periodic coordinate with period $2\pi$. The four coordinates $y^\mu$, with $\mu=0,\ldots 3$, are non-compact with the Lorentzian norm $y^2= y^\mu y^\nu\eta_{\mu\nu}$ is between -1 and 1. Locally, the spacetime is anti-de Sitter with a periodic identification of the $\chi$ coordinate,
\EQ
{
\chi\sim \chi+2\pi.
}
The conformal boundary of the spacetime is approached as $y^2\rightarrow 1$, and it is $dS_3 \times S^1$. The boundary conformal field theory is therefore formulated on a three dimensional de Sitter space with radius of curvature $R_{\rm AdS}$ times a spatial circle of radius $r_\x$. 

The geometry has a horizon at $y^2=0$, which is the three dimensional hypercone,  
\EQ
{y_0^2= y_1^2+y_2^2+y_3^2,}
and a singularity at $y^2=-1$. The singularity appears because the region where the Killing vector has negative norm needs to be excised from the physical spacetime to eliminate closed timelike curves. The hyperboloid $y^2=-1$ is a singularity since timelike geodesics end there and the Killing vector $\partial_\chi$ generating the orbifold identification has vanishing norm at $y^2=-1$. The topology of the spacetime is ${\mathbb R}^{3,1}\times S^1$, in contrast to that of the  AdS-Schwarzschild black hole which has the topology ${\mathbb R}^{1,1}\times S^3$. For this reason, in this geometry, infinity is connected, unlike in the usual Schwarzschild black hole which has two disconnected asymptotic regions.

\begin{figure}[h]
\begin{center}
\includegraphics[width=3.0in]{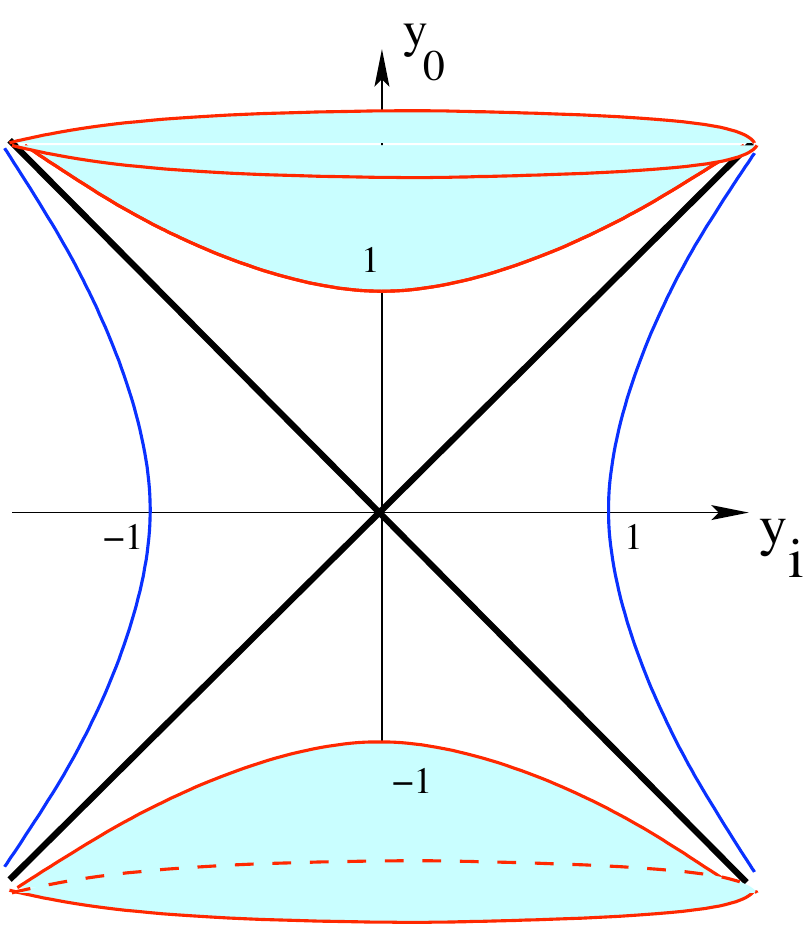} 
\end{center}
\caption{\footnotesize The global structure of the topological AdS black hole spacetime. The singularity is the hyperboloid $-y_0^2+y_i y_i=-1$ and the horizon is at the cone $y_0^2 = y_i y_i$.}
\label{spacetime}
\end{figure}

It is possible to rewrite the metric in Schwarzschild-like coordinates by introducing the following coordinate transformations
\EQ
{
Y^2=\sum_{i=1}^3y_iy_i\,;\qquad {Y\over y_0}=
\coth\left({t\over R_{\rm AdS}}\right)\,;
\qquad {r^2\over R^2_{\rm AdS}} = 4 {(Y^2-y_0^2)\over (1+y_0^2-Y^2)^2}.
}
These coordinates only  cover the exterior of the topological black hole $y^2\geq 0$. Locally, the metric takes the form
\SP
{
ds^2= {R^2_{\rm AdS} \, dr^2\over (r^2+R^2_{\rm AdS})} + 
\left({r_\x\over R_{\rm AdS}}\right)^2
(r^2+R^2_{\rm AdS})d\chi^2+
r^2
\left(-{dt^2\over \R^2}+\cosh^2 \left({t\over R_{\rm AdS}}\right)
  \;d\Omega_2^2\right). 
\label{schw}
}
The Euclidean continuation of this metric yields thermal AdS space due to the periodicity of the $\chi$ coordinate. Hence, the metric of  the exterior region of the topological black hole can also be obtained following a double Wick rotation of global AdS spacetime and a periodic identification of the  $\chi$ coordinate. In the Schwarzschild-like coordinates, the horizon of the topological black hole is at $r=0$, while the boundary is at $r \rightarrow \infty$. It is clear that each slice of constant $r$ is a $dS_3\times S^1$ geometry, while the three-dimensional de Sitter space is in the global patch. The metric (\ref{schw}),  while locally describing AdS space, differs from it globally due to the identification ${\chi\sim\chi + 2\pi}$. We note also that the spatial $S^1$ remains finite sized at the horizon, with radius  $R_{S^1}= r_\x$.

It will also be convenient to introduce a dimensionless radial coordinate $z=R_{\rm AdS}/r$, in which the metric for the exterior region is now given by
\EQ
{
ds^2= R^2_{\rm AdS} \left[{ dz^2\over z^2 (z^2+1)} +
{1 \over z^2}
\left(-{dt^2\over \R^2}+\cosh^2 \left({t\over R_{\rm AdS}}\right)
  \;d\Omega_2^2\right) + \frac{1+z^2}{z^2}
\left({r_\x\over R_{\rm AdS}}\right)^2 
d\chi^2 \right]. 
\label{in z}
}
Here, the boundary is at $z=0$, while the horizon is at $z \rightarrow \infty$.

If the bulk fermions have anti-periodic boundary conditions in the $\x$-direction, then the topological black hole has a semiclassical instability when
\EQ
{
r_{\x}<{R_{\rm AdS} \over 2\sqrt{2}}\, ,
}
which causes it to decay into the AdS bubble of nothing \cite{Balasubramanian:2005bg}. This instability can also be seen in the boundary field theory by considering the effective potential of the Polyakov loop of the Euclidean field theory \cite{Hollowood:2006xb}.

\section{The Symmetric Two-point Functions at Strong Coupling \label{calc}}

We would like to calculate the symmetric two-point function of a scalar operator ${\cal O}$ of strongly coupled ${\cal N}=4$ Super Yang-Mills in the ${\mathbb Z}_N$-invariant phase, living in $dS_3\times S^1$. This operator is dual to a scalar field $\Phi$ living in the topological $AdS_5$ black hole, with the mass $M$ of $\Phi$ is related to the dimension $\D$ of the operator ${\cal O}$ by
\EQ
{\D = 2 + \sqrt{4 + M^2 \, R_{\rm AdS}^2}.
}
To simplify notation, we will henceforth set $R_{\rm AdS}=1$. This also means that $R_{\rm dS}=1$, as the de Sitter radius of the boundary spacetime is equal to the AdS radius of the bulk. We will calculate the correlators, by first solving the equation of motion of $\Phi$ in the background (\ref{in z}), with the relevant boundary condition given by $\Phi(z,x)\big|_{\rm boundary}=\Phi_0(x)$. This equation of motion is given by
\EQ
{z^5 \partial_z \left({1+z^2 \over z^3} \, \partial_z \Phi \right) + {z^2 \over 1+ z^2} \, {1 \over r_{\chi}^2} \, \partial_{\chi}^2 \Phi + z^2 \, \Box_{dS_3} \Phi - M^2 \, \Phi = 0.
}
Since we are mainly interested in the vacuum structure of the field theory in de Sitter space, we will consider no $\chi$-dependence in $\Phi$. This is done for simplicity reason, but the generalization is straightforward. The solution to the equation of motion will then be of the form
\EQ
{\Phi(z,x) = \int dy \int dk^2 \, a_{k^2} \, R_{k^2}(z) \, F_{k^2}(x,y) \, \Phi_0(y), \label{ansatz}
}
such that 
\EQ
{\Box_{x \in dS_3} \, F_{k^2}(x,y) = k^2 F_{k^2}(x,y), \label{dssplit}
}
and
\EQ
{z^5 \partial_z \left({1+z^2 \over z^3} \, \partial_z R_{k^2} \right) + z^2 \, k^2 \, R_{k^2} - M^2 \,R_{k^2} = 0. \label{radialsplit}
}
The radial equation of motion has two solutions
\EQ
{R_{k^2}^- = z^{4-\D} \, _2F_1\left({3-\D \over 2}-{\sqrt{1-k^2} \over 2}, {3-\D \over 2}+{\sqrt{1-k^2} \over 2}; 3-\D; -z^2\right),
}
and
\EQ
{R_{k^2}^+ = z^{\D} \, _2F_1\left({\D-1 \over 2}-{\sqrt{1-k^2} \over 2}, {\D-1 \over 2}+{\sqrt{1-k^2} \over 2}; \D-1; -z^2\right).
}
As one goes to the boundary $z=\e \ll 1$, the two solutions behave as $R_{k^2}^- \rightarrow \e^{4-\D}$ and $R_{k^2}^+ \rightarrow \e^{\D}$. Therefore the most general solution is given by
\EQ
{\Phi(z,x) = \int dy \int dk^2 \, a_{k^2} \, {R_{k^2}^-(z) + \b_{k^2} \, R_{k^2}^+(z) \over R_{k^2}^- (\e)+ \b_{k^2} \, R_{k^2}^+(\e)}  \, F_{k^2}(x,y) \, \Phi_0(y), \label{sol1}
}
with the condition
\EQ
{\Phi_0(x) = \int dy \int dk^2 \, a_{k^2} \, F_{k^2}(x,y) \, \Phi_0(y).
}
We note that we could also add the term
\EQ
{\d \Phi(z,x) =  \int dy \int dk^2 \,  \gamma_{k^2} \, R_{k^2}(z)^+   \, F_{k^2}(x,y)
}
to the solution (\ref{sol1}) while still obeying the boundary conditions. However, as this term corresponds only to giving the boundary operator ${\cal O}$ non-trivial expectation value \cite{Balasubramanian:1998de}, we can omit it by requiring $\langle {\cal O} \rangle = 0$.

For explicitness, in the following calculation we will restrict ourselves to the scalar glueball operators 
\EQ
{{\cal O}= \Tr \, F_{\m\n} F^{\m\n} \qquad {\rm and} \qquad {\cal O} = \Tr \, F_{\m\n} \tilde{F}^{\m\n}. 
}
These operators are dual to the dilaton and the RR-axion, respectively, in the Type IIB Superstring theory on the bulk. The scalar glueball operators are of dimension 4, while both of the bulk fields are massless. The calculations for other operators will be similar.

In this case, the general solution to the radial equation of motion becomes
\EQ
{R_{k^2}(z) = R_{k^2}^- + \b_{k^2} \, R_{k^2}^+\, ,
}
with 
\EQ
{R_{k^2}^- = {1 \over \G({1 - \sqrt{1-k^2} \over 2}) \G({3 + \sqrt{1-k^2} \over 2})}\, _2F_1\left(-{1 \over 2}-{\sqrt{1-k^2} \over 2}, -{1 \over 2}+{\sqrt{1-k^2} \over 2}; 1; z^2+1\right),
}
and
\EQ
{R_{k^2}^+ = z^{4} \,\, _2F_1\left({3 \over 2}-{\sqrt{1-k^2} \over 2}, {3 \over 2}+{\sqrt{1-k^2} \over 2}; 3; -z^2\right).
}
Here, we have normalized this solution such that $R_{k^2}(0) = 1$.

Let us now turn to the part of the equation of motion that concerns the de Sitter factor of bulk spacetime (\ref{dssplit}). Since our goal is to obtain symmetric two-point functions in de Sitter invariant vacuum, $F_{k^2}(x,y)$ must be a function of $Z(x,y)$ and satisfies 
\EQ
{\left(Z^2-1\right)^{-1/2} \partial_Z\left[\left(Z^2-1\right)^{3/2} \, \partial_Z F_{k^2}\left(Z(x,y)\right)  \right] + k^2\, F_{k^2}\left(Z(x,y)\right) =0. \label{ds}
}
For $k^2\leq 1$, we can rewrite $\sqrt{1 - k^2} = \ell$ and we have a solution
\EQ
{F_{\ell}\left(Z\right) =  \left(Z^2-1\right)^{-{1 \over 4}} \, P_{\ell-{1 \over 2}}^{1 \over 2}\left(Z\right),
}
which, for non-negative $\ell \in {\mathbb Z}$, forms a set of complete orthonormal functions in the interval $Z \in [-1,1]$. This interval $Z \in [-1,1]$ corresponds to the point $x$ and $y$ having space-like separation. The orthogonality condition is given by
\EQ
{\int_{-1}^1 dZ \, P_{\ell-{1 \over 2}}^{1 \over 2}\left(Z\right) \, \left(P_{m-{1 \over 2}}^{1 \over 2}\left(Z\right) \right)^*= \d_{\ell,m}.
}

For $k^2\geq 1$, we can rewrite $\sqrt{1 - k^2} = i \n$ and the solution
\EQ
{F_{\n}\left(Z\right) =  \left(Z^2-1\right)^{-{1 \over 4}} \, P_{-{1 \over 2}+i \n}^{1 \over 2}\left(Z\right),
}
forms a set of complete orthonormal functions in the interval $Z \in [1,\infty)$ for non-negative $\n \in {\mathbb R}$. This interval $Z \in [1,\infty)$ corresponds to time-like separation between $x$ and $y$. The orthogonality condition is given by
\EQ
{\int_{1}^{\infty} dZ \, P_{-{1 \over 2}+i \n}^{1 \over 2}\left(Z\right) \, \left(P_{-{1 \over 2}+i \m}^{1 \over 2}\left(Z\right) \right)^*= \d(\n-\m).
}
Similarly, for $Z \in (-\infty,-1]$, the solution
\EQ
{F_{\n}\left(Z\right) =  \left(Z^2-1\right)^{-{1 \over 4}} \, P_{-{1 \over 2}+i \n}^{1 \over 2}\left(-Z\right),
}
forms a set of complete orthonormal functions for non-negative $\n \in {\mathbb R}$. This interval corresponds to the antipodal point of $y$ being time-like separated from $x$.

Hence, we find the solution to the equation of motion for the massless fields in the bulk to be
\SP
{\Phi(z,x) = & \int \limits_{\tiny \begin{array}{cc} {\rm spacelike}\\{\rm separation} \end{array}}dy \, \sum_{\ell \in {\mathbb Z}^+}  \bigg(R_{1-\ell^2}^-(z) + \b_{\ell} \, R_{1-\ell^2}^+ (z)\bigg) \, \sqrt{2 \over \pi} \,\,{ P_{\ell-{1 \over 2}}^{1 \over 2}\big(Z(x,y)\big) \over \sqrt[4]{{Z(x,y)}^2-1} } \,\,\, \Phi_0(y) \\
\\
& + \int \limits_{\tiny \begin{array}{cc} {\rm timelike}\\{\rm separation} \end{array}}dy \,\int_{0}^{\infty}  d\n\bigg(R_{1+\n^2}^- (z)+ \b_{\n} \, R_{1+\n^2}^+ (z)\bigg) \, \sqrt{2 \over \pi} \,\,{ P_{-{1 \over 2}+i \n}^{1 \over 2}\big(Z(x,y)\big) \over \sqrt[4]{{Z(x,y)}^2-1} } \,\,\, \Phi_0(y). 
}
Substituting this to the action, we can read the symmetric two-point function. For points $x$ and $y$ that have a time-like separation, the two-point function is given by
\EQ
{G_{{\b_{\n}}}(x,y) = {N^2 \over 2^{7/2} \, \pi^{5/2}} \,\int_{0}^{\infty}  d\n \,\, \Bigg[{z^2+1\over z^3} \,\, \partial_z \bigg(R_{1+\n^2}^- (z)+  \b_{\n} \, R_{1+\n^2}^+ (z)\bigg)\Bigg]_{z=\e} {P_{-{1 \over 2}+i \n}^{1 \over 2}\big(Z(x,y)\big) \over \sqrt[4]{{Z(x,y)}^2-1}} \, .
}
There is the ambiguity $\{\b_{\n}\}$ as for any given $\b_{\n}$, the radial wave function $R_{1+\n^2}=R_{1+\n^2}^- + \b_{\n} \, R_{1+\n^2}^+$ is normalizable at the horizon. This ambiguity implies that the vacuum ambiguity in de Sitter space persists at strong coupling. As the set $\{\b_{\n}\}$ parametrizes the two-point function and thus the vacua at strong coupling, we will dub this infinite class of vacua as $\{\b_{\n}\}$-vacua.

As in \cite{Hutasoit:2009xy}, the propagator for the Bunch-Davies vacuum can be obtained by requiring the radial wave function to oscillate as $R_{1+\n^2}\propto z^{-i \n}$ near the horizon, {\it i.e.}, no term that oscillates as $z^{i \n}$. Such radial wave function is given by
\EQ
{R^E_{1+\n^2}=R_{1+\n^2}^- + \b^E_{\n} \, R_{1+\n^2}^+\, ,
}
with
\EQ
{\b^E_{\n} = i {\pi\over 32}\;e^{{\pi\over 2}\nu}\;{(\nu^2+1)^2\over\cosh{{\pi\over 2}\nu}} \, . \label{be}
}
Therefore, the symmetric two-point function of the Bunch-Davies vacuum is
\SP
{G_{E}(x,y) = \, & {N^2 \over 2^{7/2} \, \pi^{5/2}} \,\int_{0}^{\infty}  d\n \,\, \Bigg[-\frac{{(1+\nu^2)}^2}{8} \Bigg(\psi\left({3-i\nu\over2}\right)+ \psi\left({3+i\nu\over2}\right)- i\pi \coth {\pi (\nu+i) \over 2} \Bigg) \\
& +\frac{{(1+\nu^2)}^2}{4}\;\left(\ln z -\gamma_E+1\right)\bigg|_{z \rightarrow \infty}+ \frac{(1+\nu^2){z^2}}{2}\bigg|_{z \rightarrow \infty}\,\, \Bigg] \, \,\, {P_{-{1 \over 2}+i \n}^{1 \over 2}\big(Z(x,y)\big) \over \sqrt[4]{{Z(x,y)}^2-1}} \, .
}
The divergent and scheme-dependent contact terms can be minimally subtracted away to yield the renormalized two-point function. We get
\EQ
{G_{E}(x,y) =  {N^2 \over 2^{11/2} \, \pi^{5/2}} \,\int_{0}^{\infty}  d\n \, {(1+\nu^2)}^2 \;\left[\psi\left({3 -i\nu \over2}\right)-{2i\nu  \over 1+\nu^2 }\right] \,\, {P_{-{1 \over 2}+i \n}^{1 \over 2}\big(Z(x,y)\big) \over \sqrt[4]{{Z(x,y)}^2-1}} \, .
}

Similarly, for two points $x$ and $y$ that have space-like separation, the symmetric two-point function is given by 
\EQ
{G(x,y) =  {N^2 \over 2^{11/2} \, \pi^{5/2}} \, \sum_{\ell} \, {(1-\ell^2)}^2 \;\left[\psi\left({3 + \ell \over2}\right)+{2 \ell \over1- \ell^2}\right] \,\, {P_{\ell-{1 \over 2}}^{1 \over 2}\big(Z(x,y)\big) \over \sqrt[4]{{Z(x,y)}^2-1}} \, .
}
We note that
\EQ
{G_{E}(x,y; \n) =  {N^2 \over 2^{11/2} \, \pi^{5/2}} \, {(1+\nu^2)}^2 \;\left[\psi\left({3 -i\nu \over2}\right)-{2i\nu  \over 1+\nu^2 }\right] \,\, {P_{-{1 \over 2}+i \n}^{1 \over 2}\big(Z(x,y)\big) \over \sqrt[4]{{Z(x,y)}^2-1}}  \label{BDmode}
}
is none other than the analytic continuation of 
\EQ
{G(x,y;\ell) =  {N^2 \over 2^{11/2} \, \pi^{5/2}}  \, {(1-\ell^2)}^2 \;\left[\psi\left({3 + \ell \over2}\right)+{2 \ell \over1- \ell^2}\right] \,\, {P_{\ell-{1 \over 2}}^{1 \over 2}\big(Z(x,y)\big) \over \sqrt[4]{{Z(x,y)}^2-1}} \, ,
}
where we have analytically continued the ``frequency" $\n = i \ell$ and used the fact $P_{\ell-{1 \over 2}}^{1 \over 2} (Z) = P_{-\ell-{1 \over 2}}^{1 \over 2}(Z)$. 

Unlike the case of the two-point function for time-like separated points, the two-point functions for points that have space-like separation do not have any ambiguity. This is due to the fact that in this case, there is only one linear combination of radial wave function that is normalizable at the horizon.

Going back to the case of points with time-like separation, the two-point function of a $\{\b_{\n}\}$-vacuum can be expressed in terms of the Bunch-Davies two-point function (\ref{BDmode}). First, let us denote the radial wave function that oscillates as $R_{1+\n^2}\propto z^{i \n}$ near the horizon as
\EQ
{R^{\tilde{E}}_{1+\n^2}=R_{1+\n^2}^- + \b^E_{-\n} \, R_{1+\n^2}^+\, ,
}
with $\b^E_{-\n}$ can be read from Eq. (\ref{be}). Then, we can express the parameter $\b_{\n}$ as
\EQ
{\b_{\n} = {\b_{\n}^E + \a_{\n} \, \b_{-\n}^E \over 1 + \a_{\n}}\, ,
}
such that the symmetric two-point function is now given by
\EQ
{G_{\b_{\n}}(x,y) =  \int_{0}^{\infty}  d\n \,\left({1 \over 1 + \a_{\n}} \, G_{E}(x,y; \n) + {\a_{\n} \over 1 + \a_{\n}} \, {G_{E}(x,y; \n)}^*\right)\, , \label{galpha}
}
where $G_{E}(x,y; \n)$ is given in Eq. (\ref{BDmode}). We denote this two-point function as $G_{\a_{\n}}(x,y)$, and instead of using $\{\b_{\n}\}$ to parametrize the infinite family of vacua, we will use $\{\a_{\n}\}$.

The two-point function of the $\{\a_{\n}\}$-vacuum has a singularity at $Z(x,y)=1$, {\it i.e.}, when the points are very close to each other or when they have a light-like separation. At short distance $d(x,y) \rightarrow 0$, the two-point function in an $\{\a_{\n}\}$-vacuum behaves like
\EQ
{G_{\a_{\n}}(x,y) = {N^2 \over {(2 \pi)}^3}  \, \int_{0}^{\infty}  d\n \,\left({\psi\left({3 -i\nu \over2}\right) + \a_{\n} \, \psi\left({3 +i\nu \over2}\right) \over 1 + \a_{\n}} - {1 -  \a_{\n} \over 1 +  \a_{\n}} \, {2i\nu  \over 1+\nu^2 }\right) \, \, \, {1 \over \big|d(x,y)\big|}\,.
}
Therefore, the two-point function of the $\{\a_{\n}\}$-vacuum differs from that of the Bunch-Davies vacuum by a factor
\EQ
{f_{\a_{\n}} = \frac{\int_{0}^{\infty}  d\n \,\left({\psi\left({3 -i\nu \over2}\right) + \a_{\n} \, \psi\left({3 +i\nu \over2}\right) \over 1 + \a_{\n}} - {1 -  \a_{\n} \over 1 +  \a_{\n}} \, {2i\nu  \over 1+\nu^2 }\right) }{\int_{0}^{\infty}  d\n \,\left({\psi\left({3 -i\nu \over2}\right) } - {2i\nu  \over 1+\nu^2 }\right) } \,.
}
It is tempting to identify an $\{\a_{\n}\}$-vacuum with an $\a$-vacuum of the weakly coupled theory by setting
\EQ
{f_{\a_{\n}} = {1+e^{\a+\a^*} \over 1-e^{\a+\a^*}} \, . \label{singular}
}
However, for a given $\a$, there is not a unique solution $\{\a_{\n}\}$ for this equation. Furthermore, for $\{\a'_{\n}\} \ne \{\a_{\n}\}$, but with $f_{\a'_{\n}} = f_{\a_{\n}}$, the two-point functions are not identical. Therefore, this identification is not valid.

\section{Discussions \label{discussions}}
In this paper, using gauge/gravity correspondence, we have calculated the symmetric two-point functions of the scalar glueball operators of strongly coupled ${\cal N}=4$ Super Yang-Mills living on three-dimensional de Sitter space (times a circle). 

The two-point functions for points with space-like separation consist of a sum of functions parametrized by discrete ``frequency," while the two-point functions for points with time-like separation are described by an integral over a continuous spectrum of ``frequency." This is the explanation for the subtlety involving discrete normalizable mode functions in de Sitter space found in \cite{Hutasoit:2009xy}. The statement is that to obtain the retarded propagators of the inhomogeneous perturbations, one has to include the contributions from these discrete normalizable mode functions (see for example  Eqs. (3.59) and (3.77) of Ref. \cite{Hutasoit:2009xy}), and the reason is because they correspond to contributions from points with space-like separation.

For points that are space-like separated, the two-point function is unique, but for those that have time-like separation, there is an infinite family of two-point functions $G_{\a_{\n}}(x,y)$, which is given by Eqs. (\ref{galpha}) and (\ref{BDmode}). This ambiguity arises as the consequence of the existence of infinite family of bulk radial wave functions that are normalizable at the horizon \cite{Balasubramanian:1998de}. This ambiguity implies that there is an infinite class of de Sitter invariant vacua for the strongly coupled theory on de Sitter space, which we have parametrized with a set of complex parameters $\{\a_{\n}\}$. The short-distance singularity of an $\{\a_{\n}\}$-vacuum differs from that of the Bunch-Davies vacuum by a factor $f_{\a_{\n}}$, which is defined in Eq. (\ref{singular}). As the short-distance behavior of the propagator of an $\a$-vacuum in the weakly coupled theory differs from the short-distance behavior of the propagator of the Bunch-Davies vacuum by a factor $\left(1+e^{\a+\a^*}\right)/\left(1-e^{\a+\a^*}\right)$, it is tempting to identify the $\{\a_{\n}\}$-vacuum at the strong coupling with the $\a$-vacuum at the weak coupling by setting these two factors to be equal. However, as for a given $\a$, there is not a unique set of $\{\a_{\n}\}$ that satisfies this condition, and as for $\{\a'_{\n}\} \ne \{\a_{\n}\}$, $G_{\a'_{\n}}(x,y) \ne G_{\a_{\n}}(x,y)$ even though $f_{\a'_{\n}} = f_{\a_{\n}}$, such identification is not valid. Since we do not have the tool to analyze the theory at intermediate coupling, we do not know how an $\a$-vacuum evolves into an $\{\a_{\n}\}$-vacuum as we go from the weak coupling regime to the strong coupling one. One possible explanation is that as one increases the coupling of the theory, there is a mixing between the $\a$-vacua. However, this explanation comes with a puzzle as the strong coupling scalar glueball correlators of the Bunch-Davies vacuum match the weak coupling ones \cite{Hutasoit:2009xy} and one might expect that the scalar glueball operators are protected and their correlators are not renormalized.

In Section \ref{alpha}, we have mentioned that at weak coupling, there is a widely used interpretation, in which the extra contributions in the $\a$-vacuum propagator (\ref{alphaweak}) are thought of as arising from an image source at the antipodal point. At strong coupling, however, such interpretation does not arise. There is no natural way to express the symmetric two-point function in an $\{\a_{\n}\}$-vacuum (\ref{galpha}) in terms of the antipodal points. This suggests that the interpretation using an image source at the antipodal point is not the most general language that one can use to describe the vacuum structure of field theories in de Sitter space.

Let us also briefly comment on the connection between vacuum ambiguity in the bulk and vacuum ambiguity of the field theory on the boundary. Since the bulk spacetime, which is the topological AdS black hole, is a time-dependent geometry, one expects that there is also vacuum ambiguity in the bulk. However, all bulk vacua, with the exception of the bulk Euclidean vacuum, feature extra singularities that render them unphysical \cite{Ross:2004cb}. Ref. \cite{Ross:2004cb} then suggests that this might mean that there should be no vacuum ambiguity in the strongly coupled dual field theory in the de Sitter boundary either. Here, after explicitly constructing the symmetric two-point functions of the strongly coupled boundary field theory, we do not find any extra singularities that will deem a generic $\{\a_{\n}\}$-vacuum in the boundary unphysical. We would like to emphasize that this lack of extra singularities is only at the level of the two-point functions. As the higher $n$-point functions are more sensitive to the properties of the bulk spacetime (see for example \cite{Polchinski:1999ry,Gary:2009ae} and references therein), it is certainly worth studying the higher $n$-point functions of the strongly coupled ${\cal N}=4$ Super Yang-Mills on $dS_3 \times S^1$ and see whether the higher $n$-point functions of a generic $\{\a_{\n}\}$-vacuum feature any extra singularities that will render such a vacuum unphysical.

Lastly, as the de Sitter boundary is a three-dimensional de Sitter space, let us mention an odd property of odd dimensional de Sitter space, namely that at the level of free theory, there exists a vacuum with no particle production \cite{Bousso:2001mw}. It would be interesting to understand what the fate of this vacuum is as one goes to the strong coupling regime.

\acknowledgments
We would like to thank Sean Hartnoll, Rich Holman and S. Prem Kumar for useful discussions and comments. This work is supported in part by DOE Grant No. DE-FG03-91-ER40682.

\end{document}